\documentclass[aps,reprint,amsmath,amssymb,groupedaddress]{revtex4-1}
\usepackage[pdftex]{graphicx}
\usepackage{bbm}

\begin{document}

\title{Codeword stabilized quantum codes on subsystems}

\author{Jeonghwan Shin}
\email{jhsh@korea.ac.kr}
\affiliation{School of Electrical Engineering, Korea University, Seoul, Korea}
\author{Jun Heo}
\email{Corresponding author: junheo@korea.ac.kr}
\affiliation{School of Electrical Engineering, Korea University, Seoul, Korea}
\author{Todd A. Brun}
\email{tbrun@usc.edu}
\affiliation{Communication Sciences Institute, University of Southern California,
Los Angeles, CA 90089, USA}

\date{\today}

\begin{abstract}
Codeword stabilized quantum codes provide a unified approach to constructing quantum error-correcting codes, including both additive and non-additive quantum codes.  Standard codeword stabilized quantum codes encode quantum information into subspaces.  The more general notion of encoding quantum information into a subsystem is known as an operator (or subsystem) quantum error correcting code.  Most operator codes studied to date are based in the usual stabilizer formalism.  We introduce operator quantum codes based on the codeword stabilized quantum code framework.  Based on the necessary and sufficient conditions for operator quantum error correction, we derive a error correction condition for operator codeword stabilized quantum codes.  Based on this condition, the word operators of a operator codeword stabilized quantum code are constructed from a set of classical binary errors induced by generators of the gauge group.  We use this scheme to construct examples of both additive and non-additive codes that encode quantum information into a subsystem.

\end{abstract}

\pacs{03.67.Pp 03.67.Hk}
\keywords{Quantum information, Quantum error correction, Non-additive quantum code, codeword stabilized quantum codes, operator quantum error correcting codes}
\maketitle

\section{Introduction}

Most quantum error-correcting codes that have been studied can be constructed from classical error correcting codes.  In \cite{PhysRevA.54.1098} it was shown that it is possible to construct quantum error correcting code from classical binary linear codes that satisfy certain conditions.  If a classical linear code satisfies a dual-containing constraint,  a quantum error-correcting code can be constructed from it.  The class of stabilizer codes \cite{Gottesman:1996ub} is a more general framework to construct quantum codes analogous to classical additive codes.

More recently, it was shown that non-additive quantum codes can also be constructed from classical codes using the codeword stabilized (CWS) framework \cite{Cross:2009jo}.  The CWS framework includes both additive and non-additive quantum error-correcting codes.  The starting point of a CWS code is a single stabilizer state (the ``codeword'' of the name), which is assumed to be a graph state \cite{PhysRevA.69.022316}.  Using the stabilizer generators of this graph state, all single-qubit Pauli errors acting on a codeword state can be mapped onto errors comprising only $Z$ and identity operators.  Because of this mapping, CWS codes correspond to classical codes designed to correct a particular set of binary errors.  From this associated classical code, which can correct these induced binary errors, a set of basis states that span the code space of the CWS code can be identified, which enables to code to correct the given set of Pauli errors.  The same thing can be done with a set of multi-qubit Pauli errors to produce codes with higher distances.  We can therefore think of CWS codes in standard form as being specified by a graph, whose vertices correspond to the qubits of the codeword, and a classical binary code  \cite{PhysRevA.69.022316}.

While the quantum codes described above encode quantum information into a subspace, a more general scheme is to encode quantum information into {\it subsystems}.  Operator quantum error correction (OQEC) \cite{PhysRevLett.94.180501, PhysRevLett.95.230504, PhysRevA.73.012340, Kribs:2005wv, Kribs:2005ti} provides a more general method to construct quantum codes.  This framework unifies both passive error-avoiding schemes, such as decoherence-free subspaces and noiseless subsystems, and active error correction.  In addition, OQEC shows that using active error correction, quantum information can be encoded into subsystems. 

In this paper, we introduce operator codeword stabilized (OCWS) quantum codes that encode quantum information into a subsystem.  We show that it is possible to construct both additive and non-additive quantum codes encoding information into a subsystem using the CWS framework.  OCWS codes are also specified by a graph state and a classical binary error-correcting code that can correct a set of errors induced by the word stabilizer.  In standard form, an OCWS code has a gauge group generated by the stabilizer generators of the base state (usually, a graph state, with stabilizers each containing a single $X$ and several $Z$ operators), plus some additional gauge operators that include only $Z$ and $I$ operators.  These additional gauge operators commute with each other, but anticommute with some of the stabilizer generators.  By applying gauge operators to the induced errors, the $Z$ operators located on certain qubits can be removed.  As a result, the word operators can act as the identity on those qubits where the gauge operators have a $Z$ operator.

This paper is organized as follows.  In Section~\ref{Sec:Review of OQEC and CWS codes}, we give a brief overview of operator quantum error correction and the construction of CWS codes.  In section \ref{Sec:Operator Codeword stabilized quantum code}, we give a detailed description of our framework for operator CWS codes, and present some examples.  Finally, in section \ref{Sec:Conclusions}, we conclude.

\section{Review of OQEC and CWS codes}
\label{Sec:Review of OQEC and CWS codes}

\subsection{Operator quantum error correction}

Operator quantum error correction \cite{PhysRevLett.94.180501, PhysRevLett.95.230504, PhysRevA.73.012340} is a generalized theory of quantum error correction (QEC) and provides a unified framework which combines active error correction and passive error avoiding schemes such as decoherence-free subspaces and noiseless subsystems.  In this framework, quantum information is encoded into a subsystem.  Consider a fixed partition of a system's Hilbert space:
\begin{equation*}
\mathcal{H}=(\mathcal{A} \otimes \mathcal{B}) \oplus \mathcal{K} .
\end{equation*}
Here, Hilbert space is partitioned into two subspaces, $\mathcal{K}$ and $\mathcal{A} \otimes \mathcal{B}$.  $\mathcal{A} \otimes \mathcal{B}$ is a orthogonal to $\mathcal{K}$, and separated into two subsystem by the tensor product structure. Quantum information can be encoded into subsystem $\mathcal{A}$ by preparing the information state $\rho^A$ in subsystem $\mathcal{A}$:
\begin{equation*}
\rho = \rho^A \otimes \rho^B \oplus 0^K ,
\end{equation*}
where $\rho^B$ is an any arbitrary state on the subsystem $\mathcal{B}$.  This subsystem is called the {\it noisy} or {\it gauge} subsystem; operations that affect only the gauge subsystem leave the encoded information unchanged.

Let $\mathcal{E}=\{E_a\}$ be a set of {\it error operators}.  For a standard quantum error-correcting code, the condition for $\mathcal{E}$ to be correctable on a code space $\mathcal{C}$ is
\begin{equation}
\label{Eq:NS_condition}
P_C E_a E_b P_C = \lambda_{ab}P_C\ \textrm{for all $a$, $b$} ,
\end{equation}
where $P_C$ is a projection operator onto the code space $\mathcal{C}$, $\lambda_{ab}$ is an Hermitian operator (with indices $a$ and $b$), and $E_a,E_b\in\mathcal{E}$ are any pair of error operators for $\mathcal{E}$.  Commonly, we associate such a set of error operators with a physically allowed (that is, completely positive trace-preserving) map.  In terms of the operator-sum representation, the map is
\begin{equation*}
\rho \rightarrow \mathcal{E}(\rho) = \sum_a E_a \rho E_a^\dagger,
\end{equation*}
so the error operators $\{E_a\}$ are Kraus operators of the error map.  (By a slight abuse of notation, we use $\mathcal{E}$ to denote both the map and the set of error operators that represent it.)

This correctability condition changes for operator codes.  For an OQEC code on Hilbert space $\mathcal{H}$ that encodes information into subspace $\mathcal{A}$, an error map $\mathcal{E}$ is correctable if there exists a physical map $\mathcal{R}$ on $\mathcal{H}$ that reverses the effect of $\mathcal{E}$, up to a transformation of subsystem $\mathcal{B}$.  In other words, if $\mathcal{E}$ is correctable, there exists a physical recovery map $\mathcal{R}$ such that for all $\rho^A$ and $\rho^B$,
\begin{equation}
\label{Eq:OQECCondition}
(\mathcal{R} \circ \mathcal{E})(\rho^A \otimes \rho^B) = \rho^A \otimes \rho'^B .
\end{equation}
for some $\rho'_B$.  Eq.~(\ref{Eq:OQECCondition}) can be satisfied if and only if the following condition is true:
\begin{equation}
\label{Eq:OQEC_TestableCondition}
P E_a^\dagger E_b P = I^A \otimes g_{ab}^B ,
\end{equation}
where $P=\mathbbm{1}^A \otimes \mathbbm{1}^B$ is the projector onto $\mathcal{A} \otimes \mathcal{B}$, and $g_{ab}^B$ is an arbitrary operator in $\mathbb{B}(\mathcal{B})$, for all $E_a, E_b \in \mathcal{E}$.

It is possible to extend the stabilizer formalism to include OQEC codes \cite{PhysRevLett.95.230504}. In this case, we encode a state of $k$ logical qubits into $n$ physical qubits.  Let $\mathcal{P}_n$ be the $n$-fold Pauli group.  The initial state before encoding can be represented by
\begin{equation}
|C\rangle=|0\rangle^{\otimes s}|\psi\rangle |\phi\rangle ,
\label{OQEC_canonical}
\end{equation}
where $|\phi\rangle$ is the $k$-qubit state we wish to encode into a subsystem, $|\psi\rangle$ is an arbitrary $r$-qubit state (which will correspond to the gauge subsystem), and the remaining $s=n-k-r$ qubits are ancillas in the state $|0\rangle$.  Even if $|C\rangle$ and $|C'\rangle=|0\rangle^{\otimes s}|\psi'\rangle |\phi\rangle$ are different (because $|\psi\rangle\not=|\psi'\rangle$), both states are considered to encode the same information in the OQEC theory.
Therefore, $|C\rangle$ and $|C'\rangle$ are equivalent by a {\it gauge transformation}:
\begin{equation*}
|C\rangle=g|C'\rangle
\end{equation*}
where $g$ is an operator in the algebra generated by the {\it gauge group} $\mathcal{G}$.

The gauge group $\mathcal{G}$ of this OQEC code is a nonabelian subgroup of $\mathcal{P}_n$ generated by
\begin{equation*}
Z_{1},\dots,Z_{s+r},X_{s+1},\dots,X_{s+r}.
\end{equation*}
Defined in this way, the gauge group includes the stabilizer group of this code, $\mathcal{S}$, that is generated by $Z_1,\dots,Z_s$.

The algebraic structure of this trivial code---that is, the set of unencoded states (\ref{OQEC_canonical})---carries over to the OQEC after encoding.  The initial state is encoded by a unitary operator $U$ in the Clifford group.  After encoding, the generators of the gauge group are $\{S_1,\dots,S_s+r,g_{s+1},\dots,g_{s+r} \}$, where $S_i$ and $g_j$ are isomorphic to $Z_i$ and $X_j$ on the unencoded state:
\begin{equation}
S_i = U Z_i U^\dagger ,\ \ \ \ 
g_j = U X_j U^\dagger .
\end{equation}
With this definition of the gauge group, the error map $\mathcal{E}$ is correctable if and only if
\begin{equation}
E_a E_b \notin N(\mathcal{S})-\mathcal{G}
\end{equation}
for all $E_a, E_b\in\mathcal{E}$ \cite{PhysRevLett.95.230504}.  We characterize an operator code by the parameters $n$, $k$, and $d$ (just as for a standard stabilizer code), but also the number of gauge qubits $r$; we write this as $[[n,k,r,d]]$.


\subsection{Codeword stabilized quantum codes}

Codeword stabilized (CWS) codes \cite{Cross:2009jo} are a broad class of quantum error-correcting codes including both additive and non-additive quantum codes, and including stabilizer codes as a subset.  CWS codes in standard form can be specified by a graph $G$ and a classical binary code (which is in general not additive).  The $n$ vertices of the graph $G$ correspond to the $n$ qubits of the code, and its adjacency matrix is $A$.  Given the graph state and the binary code, a unique base state $|S\rangle$ and a set of word operators $\{w_l\}$ are specified.

The base state is a single stabilizer state, stabilized by a maximal Abelian subgroup $\mathcal{S}$ of $\mathcal{P}_n$. We call $\mathcal{S}$ the {\it word stabilizer}.  In standard form, this word stabilizer is generated by a set of Pauli operators with the following structure:
\begin{equation}
S_i=X_iZ^{\mathbf{r}_i} ,
\end{equation}
where $S_i \in \mathcal{S}$ and $\mathbf{r}_i$ is the $i$th row vector of the adjacency matrix $A$.  We are using the shorthand notation
\[
Z^{\mathbf{v}} = Z^{v_1}\otimes Z^{v_2} \otimes \cdots \otimes Z^{v_n} ,
\]
where $\mathbf{v} = (v_1\, v_2\, \cdots\, v_n)$ is a binary $n$-vector.
  
We see that for a CWS code in standard form, the base state $|S\rangle$ is a graph state \cite{PhysRevA.69.022316}.  The code space of a CWS code is spanned by a set of basis vectors which result from applying the word operators to the base state:
\begin{equation}
|w_l\rangle=w_l|S\rangle.
\end{equation}
Therefore, the dimension of the code space is equal to the number of word operators $\{w_l\}$.  The word operators are Pauli operators in $\mathcal{P}_n$ that anticommute with one or more of the stabilizer generators for the base state.  They therefore map the base state onto an orthogonal state.  (The exception is that they generally include the identity, so that the base state is also a codeword of the quantum code.)  The span of all these basis states is the code space.  These basis states are also eigenstates of the stabilizer generators, but with some of the eigenvalues differing from $+1$.

Non-additive codes have a different notation for quantum codes.  Unlike stabilizer codes, the dimension of the code space need not be a power of 2.  We denote a quantum code that encodes a $K$-dimensional code space into $n$ physical qubits with minimum distance $d$ as an $((n,K,d))$ code.  So $[[n,k,d]]$ corresponds to $((n,2^k,d))$.

We would now like to find an appropriate condition within the CWS framework for a set of errors $\mathcal{E}$ to be correctable.  For simplicity of analysis, we will now assume that the error operators $E_a\in\mathcal{E}$ are themselves Pauli operators.  In that case, the necessary and sufficient condition for correctability is to have
\begin{equation}
\label{Eq:CWS_ECC_Condition}
\forall i \not= j,\, a \phantom{1} w_i^\dagger E_a w_j \notin S,
\end{equation}
up to an overall phase.

An important feature pointed out in \cite{Cross:2009jo} is that any error in a correctable set of errors acting on a codeword of a CWS code in standard form can be represented (up to a phase) by another error consisting only of $Z$ and identity operators, called the {\it induced error}.  This equivalent error set is found by multiplying each error operator by elements of the word stabilizer to cancel out all components of $X$.  The set of induced errors gives rise to a mapping between the set of quantum errors and a set of classical binary errors (generally acting on multiple bits).  The mapping between a Pauli error $E = Z^{\mathbf{v}}X^{\mathbf{u}}$ and a classical binary error is defined by
\begin{eqnarray}
\label{eq:CWS_map}
Cl_{G}(E = Z^{\mathbf{v}}X^{\mathbf{u}})= \mathbf{v}\oplus\bigoplus_{l=1}^nu_l\mathbf{r}_l,
\end{eqnarray}
where $\mathbf{r}_l$ is the $l$th row of the adjacency matrix for ${G}$, and $u_l$ is the $l$th bit of the vector $\mathbf{u}$.  (Operators that differ only by a phase are mapped to the same bit string.)  Using this definition, Theorem 3 of \cite{Cross:2009jo} states that a CWS code in standard form, characterized by a graph $G$ and a classical binary code $C_b$, detects errors from a set $\mathcal{E}$ if and only if $C_b$ detects errors from the set $\{Cl_{G}(E_a)\}$ for all $E_a\in\mathcal{E}$, and if for each $E_a\in\mathcal{E}$,
\begin{eqnarray}
\textrm{either}\phantom{1}Cl_{G}(E_a) &\neq& 0 , \\
\textrm{or, for each}\phantom{1}l,\phantom{1}Z^{\mathbf{c}_l} E_a&=&E_a Z^{\mathbf{c}_l} ,
\end{eqnarray}
where the $\mathbf{c}_l$ are the codewords from the classical binary code $C_b$.  So we see that the word operators $w_l$ of the CWS code are derived from the codewords of the binary code by
\begin{equation}
\mathcal{W} = \{ w_l \} = \{Z^{\mathbf{c}_l}\}_{\mathbf{c}_l\in\mathcal{C}_b} .
\end{equation} 

\section{Operator Codeword Stabilized Quantum Codes}
\label{Sec:Operator Codeword stabilized quantum code}

In this section, we generalize the framework for CWS codes to encode quantum information into a subsystem.  Such operator CWS (OCWS) codes are defined by a gauge group and set of word operators, similar to standard CWS codes that are specified by a word stabilizer and set of word operators.  (As above, we define the gauge group to include the stabilizer operators.)

\subsection{Base state of the canonical code}

Just as in our discussion of operator codes above, for simplicity we first consider the initial (unencoded) base state $|S'\rangle$ of the OCWS code, which consists of $s=n-r$ qubits in the state $|0\rangle$ and $r$ gauge qubits in an arbitrary state:
\begin{equation}
\label{Eq:InitialState}
|S'\rangle=|0\rangle^{\otimes s}|\psi\rangle
\end{equation}
where $|\psi\rangle$ is an arbitrary $r$-qubit state.
In density operator form, Eq.~(\ref{Eq:InitialState}) is
\begin{equation*}
\rho=|S'\rangle\langle S'|=\rho^A \otimes \rho^B
\end{equation*}
where $\rho^A=(|0\rangle\langle 0|)^{\otimes s}$ and $\rho^B=|\psi\rangle\langle \psi|$.  Two base states with different states $|\psi\rangle$ and $|\psi'\rangle$ are considered equivalent.

This equivalence class of base states has a stabilizer group $\mathcal{S}$, i.e.,
\begin{equation*}
s|S'\rangle=|S'\rangle,\phantom{1} \textrm{for all $s \in \mathcal{S}$}.
\end{equation*}
The base state consists of a fixed $s$-qubit state in subsystem $\mathcal{A}$ and an arbitrary $r$-qubit state in subsystem $\mathcal{B}$.  To stabilize this base state, the stabilizer group must be a maximal Abelian group that stabilizes the state of subsystem $\mathcal{A}$ while acting as the identity on subsystem $\mathcal{B}$.   For our canonical code (\ref{Eq:InitialState}), the fixed state is $|0\rangle^{\otimes s}$, and the stabilizer group is generated by the operators
\begin{equation*}
\mathcal{S}=\langle Z_1,\cdots,Z_{s}\rangle.
\end{equation*}

The gauge group $\mathcal{G}$ of the base state is generated by stabilizer generators and operators acting only on subsystem $\mathcal{B}$.  Therefore, for our canonical code (\ref{Eq:InitialState}) the gauge group $\mathcal{G}$ of the base state is generated by
\begin{eqnarray*}
\mathcal{G}=\langle Z_1,\dots,Z_s,Z_{s+1},\dots,Z_n,X_{s+1},\dots,X_n \rangle .
\end{eqnarray*}
Gauge operators act trivially on subsystem $\mathcal{A}$, and leave the equivalence class of base states invariant:
\begin{eqnarray*}
\left\{\begin{array}{l}
g|S'\rangle = |S''\rangle\\
\rho_A = \mathrm{Tr}_\mathcal{B} \{ |S'\rangle\langle S'| \}
 = \mathrm{Tr}_\mathcal{B} \{ |S''\rangle\langle S''| \} ,
\end{array}\right.
,\phantom{1} \forall g \in \mathcal{G}.
\end{eqnarray*}

\subsection{Word operators}

In a standard CWS code one produces the basis states of the code by applying word operators to the base state.  We will retain a similar structure for OCWS codes.  By applying word operators $w_l$ to the base state, we produce a set of codewords:
\begin{equation}
\label{Eq:WordOperatorState}
w_l|S'\rangle=|w_l'\rangle.
\end{equation}
Since these codewords must all be distinct and orthogonal, at most one word operator can be in gauge group $\mathcal{G}$.  We must also choose the word operators so that no information about subsystem $\mathcal{A}$ can leak into subsystem $\mathcal{B}$.  A natural way to do this is for the word operators to act nontrivially only on subsystem $\mathcal{A}$.

For the canonical code (\ref{Eq:InitialState}), all $Z$ operators in $\mathcal{P}_n$ are in the gauge group; and the $X$ operators on $\mathcal{B}$ are in $\mathcal{G}$ as well.  The word operators must be elements of $\mathcal{W} = \mathcal{P}_n/\mathcal{G} = \langle X_1,\dots,X_s \rangle$, i.e., the $X$ operators acting on subsystem $\mathcal{A}$.  We choose $K$ linearly independent word operators from this group $\mathcal{W}$.  (Generally we include the identity as one of the word operators so that the base state is also a codeword.)

For the initial base state in Eq.~(\ref{Eq:InitialState}), the word operators have the form of
\begin{equation*}
w_l \equiv w_l^A\otimes I^B ,
\end{equation*}
where $I^B$ is trivial operator on $\mathcal{B}$.  More generally, the word operators could take the form $w_l^A \otimes g$ for some $g\in\mathbb{B}(\mathcal{B})$, having a non-trivial operator $g$ acting on $\mathcal{B}$.
In this case, Eq.(\ref{Eq:WordOperatorState}) becomes
\begin{eqnarray*}
(w_l^A \otimes g) |S'\rangle &=& w_l^A|0\rangle^{\otimes s}\otimes g|\psi\rangle\\
&=&w_l^A|0\rangle^{\otimes s} \otimes |\psi'\rangle .
\end{eqnarray*}
However, clearly we must use the same operator $g$ for every word operator $w_l$, or we introduce a correlation between $\mathcal{A}$ and $\mathcal{B}$ that violates the operator code structure.  Since such an operator $g$ can be absorbed into the encoding unitary $U$ (see below), without loss of generality we assume the word operators take the form $w_l^A \otimes I^B$.

\subsection{Encoding and the error-correcting condition}

From the canonical code described above we derive the properties of a general OCWS code.  We apply an encoding unitary $U$ to the codeword states above.  The base state (\ref{Eq:InitialState}) becomes $|S'\rangle \rightarrow U(|0\rangle^{\otimes s} \otimes |\psi\rangle)$.  The gauge group transforms to
\[
\mathcal{G} = \langle UZ_1U^\dagger ,\ldots, UZ_nU^\dagger, UX_{s+1}U^\dagger, \ldots, UX_nU^\dagger \rangle ,
\]
and the word operators (\ref{Eq:WordOperatorState}) transform as $w_l \rightarrow Uw_lU^\dagger$.

We can derive the error correction conditions for OCWS codes from Eq.~(\ref{Eq:OQEC_TestableCondition}).  To {\it detect} if an error has occurred, it is necessary and sufficient to satisfy
\begin{equation*}
\left(\langle w_i^A|\langle k|\right) E \left(|w_j^A\rangle |l\rangle\right) = c_{Ekl}\delta_{ij}
\end{equation*}
for all $E \in \mathcal{E}$, where $|w_{i,j}^A\rangle$ are the components of the codeword states $|w_{i,j}'\rangle$ on subsystem $\mathcal{A}$, and $|k\rangle$ and $|l\rangle$ are basis states on subsystem $\mathcal{B}$.
By Eq.~(\ref{Eq:WordOperatorState}), we see that
\begin{equation*}
\langle S'|w_i E w_j|S'\rangle=c_E\delta_{ij}
\end{equation*}
where $c_E$ depends on $E$ and the qubit state $|\psi\rangle$ on $\mathcal{B}$.  Since the word operators transform eigenvectors of $\mathcal{G}$ to other eigenvectors, we see that
\begin{eqnarray*}
w_l g |S'\rangle &=& w_l |S''\rangle = |w_l'' \rangle , \\
g w_l |S'\rangle &=& \pm w_l g |S'\rangle = \pm|w_l''\rangle ,
\end{eqnarray*}
for any $g \in \mathcal{G}$.  This leads to necessary and sufficient conditions to detect errors:
\begin{equation}
\label{Eq:OCWSCondition}
\forall i \not= j,\, g \in \mathcal{G},  \phantom{1} w_i E w_j \ne g ,
\end{equation}
up to an overall phase.



Given the ability to detect errors, we can correct them with the same binary code construction used in standard CWS codes.  In standard form, the base state is a graph state.  After unitary encoding, the $Z$ and $X$ operators can be mapped by the unitary encoding operator $U$ as follows:
\begin{eqnarray*}
U:\left\{\begin{array}{l}Z_i \rightarrow X_i Z^{\mathbf{r}_i} \\
X_i \rightarrow Z_i\end{array}\right.
\end{eqnarray*}
where $\mathbf{r}_i$ is the $i$th row vector of the adjacency matrix of the graph.  Therefore, the gauge group of the OCWS code can be generated by
\begin{equation*}
\mathcal{G}=\langle \{S_i, g_j\}\rangle
\end{equation*}
where
\begin{eqnarray*}
\begin{array}{ll}
S_i = X_i Z^{\mathbf{r}_i}, & \textrm{for $i=1,\dots,n$} \\
g_j = Z_{s+j}, & \textrm{for $j=1,\dots,r$.}
\end{array}
\end{eqnarray*}
One of the elegant features of CWS codes in standard form is that all errors can be represented by operators consisting only of $Z$s and identity operators.  Any $X$ or $Y$ operators can be mapped onto $Z$s by multiplying them by the appropriate generator $S_i$.

For OCWS codes, a further reduction is possible.  The $Z$ operators in an induced error located at qubits $s+1,\dots,n$ can also be removed by being multiplied by $g_j$ operators.  This can map multiple errors onto the same equivalent binary error; this is a common feature of operator codes, because the errors differ by a pure gauge operator that affects only the noisy subsystem.  We can then map each error in a set to a classical binary error, with $0$s for each $I$ operator and $1$s for each $Z$ operator.

Generically, the word operators $\{w_l\}$ can also be taken to contain only $Z$ and $I$ operators.  We must be careful, because the word operators must be chosen to all act identically on the gauge subsystem, as described above.  If we take the natural convention that the word operators act on the gauge subsystems as the identity, this requires that each $w_l$ commutes with the gauge subgroup.  For the standard form where the base state is a graph state, this means that the $\{w_l\}$ must commute with the gauge operators $g_1,\ldots,g_r$ and their anticommuting partners $S_{s+1},\ldots,S_n$.  This means that the $\{w_l\}$ operators act as the identity on qubits $s+1,\ldots,n$.

We then get the condition for a set of errors to be correctable by a given OCWS code.  The set of errors must be detectable, as summarized in the condition (\ref{Eq:OCWSCondition}) above, and the equivalent set of binary errors must be correctable by a classical binary code whose codewords correspond to the word operators $\{w_l\}$.  We illustrate this principle with some examples in the next section.

\subsection{Examples}

In this section, we give examples of OCWS codes.  All these codes use a base state based on the ring graph.

\subsubsection{Ring graphs}

Straightforwardly enough, a ring graph consists of $n$ vertices arranged in a closed loop, so each vertex has exactly two neighbors.  We will use this structure to define the base states of our OCWS codes.  For example, suppose an OCWS code has $n=5$ and $r=2$.  The gauge group of this code is generated by
\begin{eqnarray*}
S_1 &=& XZIIZ\\
S_2 &=& ZXZII\\
S_3 &=& IZXZI\\
S_4 &=& IIZXZ\\
S_5 &=& ZIIZX\\
g_1 &=& IIIZI\\
g_2 &=& IIIIZ .
\end{eqnarray*}
We can use this group to map all single-qubit errors onto induced errors containing only $Z$ and $I$ operators.  First, we apply the stabilizer elements of the graph state.  By applying $S_i$, all possible $X$, $Y$ and $Z$ errors can be represented as induced errors:
\begin{eqnarray*}
\begin{array}{cccccc}
Z : & ZIIII & IZIII & IIZII & IIIZI & IIIIZ \\
X : & IZIIZ & ZIZII & IZIZI & IIZIZ & ZIIZI \\
Y : & ZZIIZ & ZZZII & IZZZI & IIZZZ & ZIIZZ \end{array}
\end{eqnarray*}
In addition, applying $g_i$ to the induced error, $Z$ operators located on the $4$th and $5$th qubits can be removed:
\begin{eqnarray*}
\begin{array}{cccccc}
Z : & ZIIII & IZIII & IIZII & IIIII & IIIII \\
X : & IZIII & ZIZII & IZIII & IIZII & ZIIII \\
Y : & ZZIII & ZZZII & IZZII & IIZII & ZIIII \end{array}
\end{eqnarray*}

All of the examples in the rest of this section are based on ring graphs; the particular choices of word operators and binary codes were found by numerical search.

\subsubsection{$[[8,1,1,3]]$ OCWS code}

A $[[8,1,1,3]]$ code can be constructed using OCWS framework.
The gauge group of $[[8,1,1,3]]$ OCWS code is generated by
\begin{eqnarray*}
S_i &=& ZXZIIIII \phantom{1} \textrm{and cyclic shifts},\\
g_1 &=& IIIIIIIZ
\end{eqnarray*}
There are $21$ induced single-qubit Pauli errors on this $8$-qubit  code after applying the word stabilizer generators and gauge operators by Eq.~(\ref{eq:CWS_map}).  The classical binary code that can correct these induced binary errors has codewords
\begin{equation}
\label{Eq:ExErr1}
00000000 \phantom{1} 01100110.
\end{equation}
From Eq.~(\ref{Eq:ExErr1}), the word operators of this code are $w_l=Z^{\mathbf{c}_l}$:
\begin{equation*}
IIIIIIII \phantom{1} IZZIIZZI
\end{equation*}

\subsubsection{[[9,3,1,3]] OCWS code}

Using nine physical qubits, we can construct an OCWS code protecting three logical qubits from all single-qubit Pauli errors.  Based on the single ring graph state of length 9, the gauge group is generated by
\begin{eqnarray*}
S_i &=& ZXZIIIIII \phantom{1} \textrm{and cyclic shifts},\\
g_1 &=& IIIIIIIIZ
\end{eqnarray*}
Using the elements of the gauge group, the 27 single-qubit Pauli errors can be represented by $24$ classical binary errors. The word operators corresponding to classical binary codewords that can detect and correct these errors are
\begin{eqnarray*}
\left.\begin{array}{cccc}
IIIIIIIII & IZIIZZIZI & IZZZZZIII & ZIIZIZZZI \\
ZIZIIZZII & ZZIZZIZII & ZZZIZIZZI & IIZZIIIZI
\end{array}\right.
\end{eqnarray*}

\subsubsection{((9,4,1,3)) OCWS code}

Using the same base state and gauge operators as the $[[9,3,1,3]]$ OCWS code above, we can also construct a $((9,4,1,3))$ non-additive code using the OCWS framework.  The $((9,4,1,3))$ non-additive code has the same gauge group as the $[[9,3,1,3]]$ OCWS code, but the word operators are
\begin{eqnarray*}
\left.\begin{array}{cccc}
IIIIIIIII & IZIIIZZII & ZIIIZZIZI & ZIZZIIZZI
\end{array}\right.
\end{eqnarray*}

Note that all of the examples presented here were of distance 3 codes, and thus able to correct single-qubit errors.  There is nothing special about distance 3, however; one can similarly design OCWS codes to correct larger numbers of errors in exactly the same way.

\section{Conclusions}
\label{Sec:Conclusions}

We have presented a scheme to construct operator quantum error-correcting codes based on the codeword stabilized (CWS) quantum code framework.  From the necessary and sufficient conditions for error detection and correction in OQEC codes, we derived error detection and correction conditions for OCWS codes.  Using this condition, we showed that the word operators of an OCWS code can be constructed from a classical binary code designed to correct a set of binary errors representing the correctable set of quantum errors by applying the stabilizer and gauge operators.  Just as in the standard form of CWS codes, we choose the base state of an OCWS code to be a graph state; in that case, we can add gauge operators consisting only of $Z$ operators to the gauge group (which also includes the stabilizer operators of the base state). Applying these gauge operators to the set of ``induced error'' operators containing only $Z$ and $I$ operators, allows us to reduce the number of $Z$ operators in these operators. We presented several example codes based on a single ring topology, including both additive and non-additive codes.

\begin{acknowledgments}
This research was supported by the KCC(Korea Communications Commission), Korea, under the R\&D program supervised by the KCA(Korea Communications Agency)(KCA-2012-12-911-04-003).
TAB was supported in part by NSF Grant CCF-0830801.
\end{acknowledgments}

\bibliography{Operator_CWS}

\end{document}